\documentclass[10pt]{article}
\usepackage{graphics, epsfig,colordvi,afterpage}
\usepackage{times}
\usepackage{multirow}
\usepackage{palatino}
\usepackage{amsmath,amssymb}
\usepackage{caption}
\usepackage{placeins}
\usepackage{etoolbox}
\def\AnnotatedKeys{Zhaoping_iScience2025}

\makeatletter
\pretocmd{\@bibitem}{\def\currentbibkey{#1}}{}{}
\pretocmd{\@lbibitem}{\def\currentbibkey{#2}}{}{}
\makeatother

\newcommand{\printannotation}{%
  \ifinlist{\currentbibkey}{\AnnotatedKeys}{%
    \par\smallskip\small\itshape
    \csname annotation@\currentbibkey\endcsname
  }{}%
}

\makeatletter
\apptocmd{\end@bibitem}{\printannotation}{}{}
\makeatother

\usepackage{wrapfig}
\usepackage{xcolor}

\mathchardef\mhyphen="2D
\input{epsf}

\def\be{\begin{equation}}
\def\ee{\end{equation}}
\def\bea{\begin{eqnarray}}
\def\eea{\end{eqnarray}}

\def\bibsinglebull{$\phantom{\bullet}\bullet$~}
\def\bibdoublebull{$\bullet\bullet$~}

\makeatletter
\renewenvironment{thebibliography}[1]
     {\section*{\refname}%
      \@mkboth{\MakeUppercase\refname}{\MakeUppercase\refname}%
      \list{%
      \phantom{$\bullet$~}%
      \ifnum\c@enumiv=13\bibdoublebull\else     
      \ifnum\c@enumiv=36\bibdoublebull\else     
      \ifnum\c@enumiv=6\bibsinglebull\else     
      \ifnum\c@enumiv=33\bibdoublebull\else   
      \ifnum\c@enumiv=51\bibsinglebull\else   
      \ifnum\c@enumiv=61\bibdoublebull\else   
      \ifnum\c@enumiv=35\bibsinglebull\else   
    \phantom{$\bullet\bullet$~}\fi\fi\fi\fi\fi\fi\fi
      \@biblabel{\@arabic\c@enumiv}}%
           {\settowidth\labelwidth{\@biblabel{#1}}%
            \leftmargin\labelwidth
            \advance\leftmargin\labelsep
            \@openbib@code
            \usecounter{enumiv}%
            \let\p@enumiv\@empty
            \renewcommand\theenumiv{\@arabic\c@enumiv}}%
      \sloppy
      \clubpenalty4000
      \@clubpenalty \clubpenalty
      \widowpenalty4000%
      \sfcode`\.\@m}
     {\def\@noitemerr
       {\@latex@warning{Empty `thebibliography' environment}}%
      \endlist}
\makeatother

\begin{document}
\baselineskip = 14 pt
\footskip = 0.75 in
\setlength{\parindent}{ 0.33 in}

{\Large \bf Vision as looking and seeing through a bottleneck}

\centerline{Li Zhaoping}

\centerline{University of T\"ubingen, Max Planck Institute for Biological Cybernetics, T\"ubingen, Germany}
\centerline{email: li.zhaoping@tuebingen.mpg.de}

\centerline{in press for {\it Current Opinion in Neurobiology}, 2026}

%
%
%
%
%


%
%
%

\subsection*{Abstract:} 

Progress in vision research has been slower downstream than upstream of primary visual cortex (V1).
Traditional frameworks have largely overlooked a central constraint:
only a tiny fraction of retinal input is recognized.  Thus, to a first approximation, vision is
better formulated as looking and seeing through a bottleneck.  Looking, mainly by the
peripheral visual field, selects visual information to enter this bottleneck, 
largely via gaze shifts that center selected contents at fovea.  
Seeing, mainly by the central visual field, recognizes this content.
Converging evidence suggests that V1 initiates the bottleneck and contributes
to looking by generating a bottom-up saliency map that guides saccades exogenously, and that
top-down feedback along the visual pathway, targeting mainly the representation of the central visual field,
refines seeing.  Progress will accelerate through falsifiable theories that explicitly link
behavior with neural substrates, and by experimental designs that avoid forced fixation and
precisely track gaze.

\section*{Highlights} 

\begin{itemize}
\item The brain's bottleneck forces vision to select content for seeing by looking
\item Traditional research frameworks have focused on seeing and overlooked looking 
\item Logically, looking can occur before seeing, as is evident behaviorally 
\item V1 serves looking, initiates the bottleneck, and receives feedback to aid seeing
\item Looking and seeing generalize to orienting and recognizing multisensorily
\end{itemize}

\section*{Keywords}
the looking-and-seeing framework, bottleneck, attention, selection, decoding, recognition, feedforward, feedback, primary visual cortex,
visual pathway

\section*{The problem of vision}

It appeared natural to assume that the purpose of vision is to see, i.e., to identify
what is where in the visual field. 
This motivated a research approach to discover what visual features 
activate individual neurons.  Such discoveries progressed within a decade
from retina \cite{Kuffler1953} to primary visual cortex (V1)  \cite{HubelWiesel1962}. 
The features that excite V1 neurons appear understandable from those that 
excite retinal neurons \cite{HubelWiesel1962}.  
However,  progress further downstream along the visual pathway has been 
slower than expected. I argue that this is largely 
due to a misformulation of vision as mainly seeing.

\begin{figure}[h!!!]
\begin{center}
\includegraphics[width=150mm]{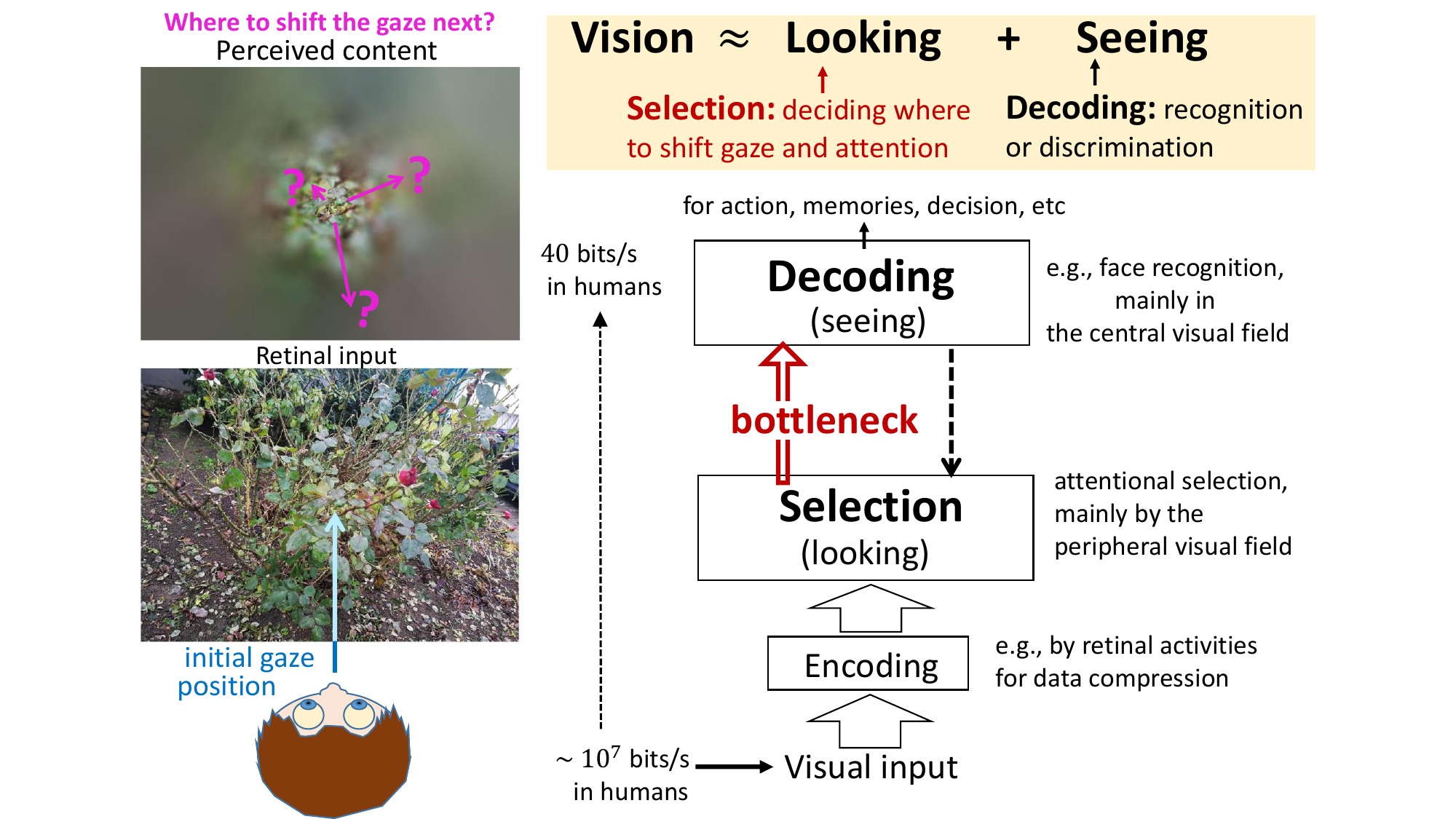}
\end{center}
\caption{\label{fig:Formulation}
A new framework to formulate vision as mainly looking and seeing through a bottleneck.
Traditional ideas view vision as mainly seeing, i.e., recognizing what is where in visual field.
However, due to a processing bottleneck, humans consciously recognize only 40 bits out of  $10^7$ bits
of retinal input information each second \cite{Sziklai1956, ZhaopingBook2014}.  
Only scene fragments around our gaze are seen clearly, due to visual crowding \cite{LeviCrowdingReview2008}.  
Hence, vision must select which input fraction to see.  
This selection is looking, largely by deciding where to shift our gaze or attention.
Humans make saccades (ballistic gaze shifts)  three times per seconds.
When saccades target destinations outside the central zone of clear visibility, 
the destinations are largely non-random \cite{LiangZhaoping2024}.  These observations and 
considerations motivate this central-peripheral dichotomy \cite{ZhaopingNewFramework2019}: 
vision in the peripheral visual field is specialized for looking, 
selecting a peripheral visual location for the next gaze or attentional shift, whereas vision in the central visual field 
is specialized for seeing.  Looking and seeing are the selection and decoding stages 
when vision is viewed as having three stages: encoding, selection, and decoding, 
with encoding as mainly to sample and efficiently represent visual inputs before selection \cite{ZhaopingBook2014}. 
}
\end{figure}

We only see, or recognize, a tiny fraction of visual inputs because the brain has 
a processing, or attentional, bottleneck \cite{SimonsChabris1999}, constrained 
by metabolic energy, space for neurons and wiring, and time for complex computations.  
For example, when we direct our gaze to (i.e., fixate on) the first letter of a word in this sentence,  
individual letters in the next word are typically illegible \cite {StrasburgerEtAl2011}.  
Likewise, in a visual scene (Figure \ref{fig:Formulation}),  only a small region around our 
gaze is clearly visible. Vision must decide where in the scene to direct the next gaze shift.
Hence, as a first approximation, vision should encompass {\it looking and seeing}:
looking selects the fraction of visual input for entry into the
bottleneck, largely by gaze shifts to center the selected content at
our fovea, and seeing recognizes the selected content.

Previous formulations that divided vision into low-, mid-, and high-level stages \cite{Palmer1999} 
are imprecise and thus difficult to test.  Marr's influential formulation notably omitted the selection (looking) 
stage by viewing vision as successively building  primal sketch,  2.5D, and 3D representations of the scene \cite{Marr1982}. 
My recent textbook \cite{ZhaopingBook2014} characterizes vision as comprising three stages: encoding, selection, 
and decoding.  Looking and seeing correspond to selection and decoding.  
Encoding --- efficient sampling and representing visual information prior to selection --- characterizes largely 
what retinal (and partly V1) neural receptive fields do \cite{ZhaopingBook2014}.

Blind to our own blindness, we have the illusion of seeing everything clearly
because our scene appears clear wherever we direct our gaze --- like assuming that the refrigerator 
light is always on because it is on whenever we open the door \cite{Thomas1999, ZhaopingPeripheral2024}.
This illusion has impeded progress.

\section*{Looking can occur before or without seeing}

\begin{figure}[h!!!]
\begin{center}
\includegraphics[width=150mm]{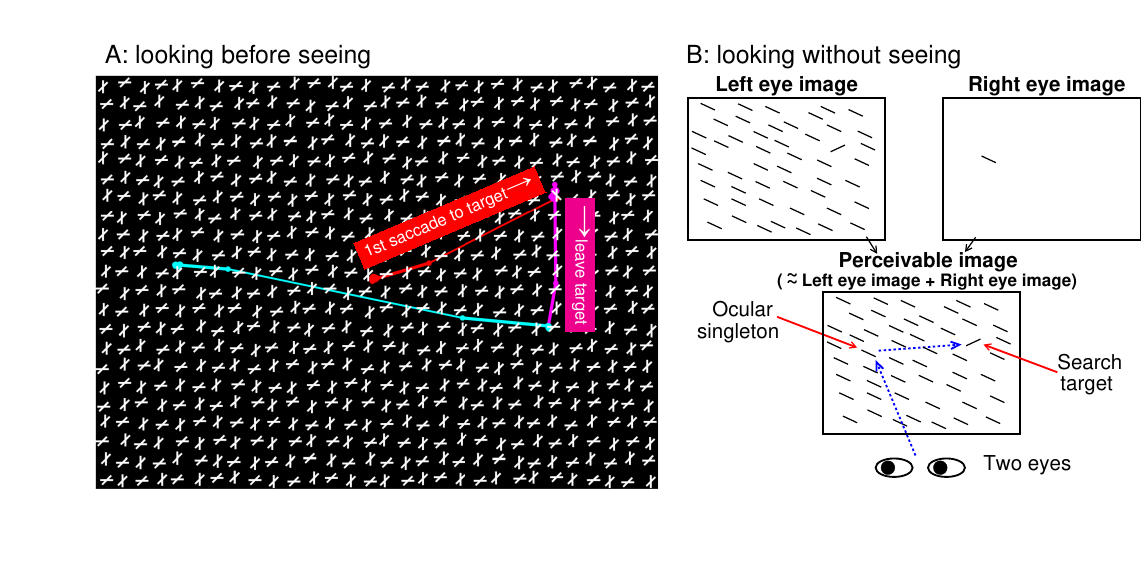}
\end{center}
\caption{\label{fig:LookingAndSeeing}
Demonstration of looking before (A) and without (B) seeing in visual search.  
In both examples, gaze started at the center of the visual search image, and observers should
search for the target  as quickly as possible. 
A: a black and white search image,  plus the initial segments of a gaze trajectory  (with annotations) of 
an observer's search for a uniquely oriented bar.
The first saccade (red) landed on the target bar, uniquely tilted counterclockwise from vertical.
Then, without crowding, foveal vision 
recognized an {\sf X} shape formed by the target bar and an intersecting 
vertical bar.  Because this {\sf X} is identical in shape (up to a rotation) to other {\sf X}'s 
in the image, the observer mistakenly vetoed the target (after the gaze stayed around the target for about 0.5 second)
and continued searching elsewhere.  
If the target bar was tilted only $20^o$ from the intersecting vertical bar to 
make the  {\sf X} uniquely thinner, the gaze departure and confusion did not occur.
B: a schematic of a trial to search for an uniquely oriented bar (tilted $50^{\circ}$ from the uniformly oriented non-target bars).
All bars except one are shown only to the left eye, a non-target ocular singleton 
is shown to the right eye.  Despite being
perceptually indistinguishable, the ocular singleton captured the first 
saccade, followed by a second saccade to the target.
The target and the ocular singleton distractor had 
equal eccentricity from the initial fixation before the search started.
When this eccentricity was sufficiently large (e.g., $12^{\circ}$),
gaze distraction in the first saccade occurred in most trials
(typically within 300 milliseconds of the appearance of the visual inputs).
In both A and B, the first saccade was directed to the most 
salient (though not necessarily distinctive perceptually) scene location, 
guided by saliency signals computed in V1.  Figure adapted from \cite{ZhaopingPeripheral2024}.
}
\end{figure}

It is logical, though unintuitive, that some looking must precede 
seeing.  Figure \ref{fig:LookingAndSeeing} illustrates this in two examples 
when observers searched for a uniquely oriented bar, which is 
salient to attract attention or gaze exogenously (reflexively) \cite {TreismanGelade80, WolfeEtAl1989, DuncanHumphreys89}.
In Figure \ref{fig:LookingAndSeeing}A, the first saccade during search 
landed on the target bar, yet this bar's unique orientation was not 
recognizable before the saccade began, 
much as text letters too far from your gaze are unrecognizable due to visual crowding \cite{LeviCrowdingReview2008}.
This looking before seeing was not accidental: in untrained observers, 
gaze reached the target within a second in 50\% of such search trials \cite{ZhaopingPeripheral2024}.
Then, recognizing the foveal {\sf X} containing the target bar triggered a confusion,
prompting the gaze to abandon the target to resume searching elsewhere. 
This search task is atypical, it can be accomplished solely by looking and is actually hindered 
by seeing, demonstrating that looking and seeing are separate rather than seamlessly integrated.

In Figure \ref{fig:LookingAndSeeing}B, gaze is distracted by a non-target bar
distinguished only by its eye of origin (of visual input), a feature invisible 
to seeing even in the fovea. This distractor frequently captures the first saccade.
Observers could not discriminate it from the other non-target bars, and were often unaware that it 
had distracted their gaze \cite{Zhaoping2008OcularSingleton, ZhaopingPeripheral2024}.

\section*{V1's pivotal function and the central-peripheral dichotomy theory }

\begin{figure}[h!!!]
\vskip -0.2 in
\begin{center}
\includegraphics[width=150mm]{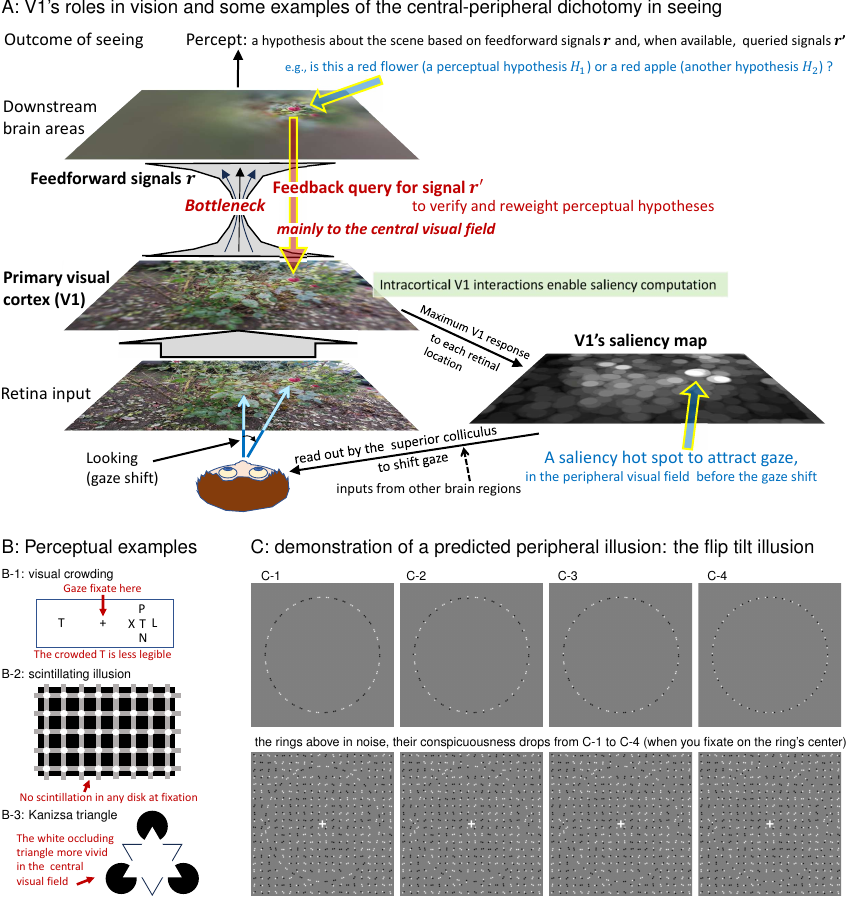}
\end{center}
\caption{\label{fig:V1Role}
\footnotesize V1's roles in vision and the central-peripheral dichotomy (CPD) in seeing.
A:  schematic of V1 functions. V1 creates a saliency map of the visual field to guide gaze 
shift exogenously(reflexively), initiates the bottleneck in information flow, 
and supplies additional information queried by 
top-down feedback from downstream stages to support ongoing perceptual processing 
under this bottleneck.  
The feedforward signals ${\bf r}$ that pass through the bottleneck and the 
feedback-queried signals
 ${\bf r'}$ (containing information absent in ${\bf r}$) 
occupy subspaces of the neural signals upstream of the bottleneck.  
The feedback is directed  mainly to the central visual field, 
producing a central-peripheral dichotomy in seeing.  
B: examples of the CPD in seeing.
Crowding and the scintillating illusion occur in the peripheral 
visual field, where feedback queries to disambiguate percepts are lacking.  
Vivid Kanizsa triangle appears in the central visual field, where analysis-by-synthesis occurs in 
the feedback query.
C: the flip tilt illusion \cite{ZhaopingFlipTilt2020}: a hetero-pair of dots (one black and one white) appears tilted 
orthogonally to its actual tilt (of the alignment between the dots) in the peripheral visual field.
This is illustrated by four rings of dot pairs.  
The homo-pairs of dots (two dots both white or both black) 
evoke no illusions, and are all parallel to the tangent of the rings. 
A ring's circumference lies in the peripheral field when you fixate at the ring's center.  
The C-4 ring is made by replacing the homo-pairs in the C-1 ring with
hetero-pairs.  
The C-1 ring appears stronger than the C-4 ring,  especially in 
noise, because hetero-pairs appear (illusorily) orthogonal to the tangent,
giving the two rings a ``snake" and ``ladder" appearance \cite{MayHess2007}.  
On the C-2 or C-3 ring, hetero-pairs are orthogonal or parallel to the tangent, respectively.
The illusion makes the C-2 ring more conspicuous than 
the C-3 ring (in noise) in the peripheral field, but
directing gaze to a segment of the ring allows
central vision to veto the illusion, making
the C-3 ring more recognizable.
}
\end{figure}

The unique  eye of origin of the distractor in Figure \ref{fig:LookingAndSeeing}B is 
visible only to V1 among all visual cortical areas, since only V1 has
a substantial percentage of neurons that are monocular (in the binocular visual field) to signal eye-of-origin.
Its ability to capture gaze was predicted by the V1 Saliency Hypothesis (V1SH), which
proposes that V1 constructs a bottom-up saliency map to guide gaze exogenously \cite{LiTICS2002} via 
the superior colliculus (Figure \ref{fig:V1Role}A).
Intracortical V1 interactions cause nearby neurons tuned to similar features,  such as orientation, color, motion direction, 
or eye-of-origin, to suppress each other \cite{RocklandLund1983, GilbertWiesel1983, AllmanEtAl1985, KnierimVanEssen1992, ZhaopingBook2014}.
Accordingly, background elements in Figure \ref{fig:LookingAndSeeing} evoke
suppressed V1 responses; meanwhile, 
the feature singletons escape iso-feature suppression,  evoking stronger responses, 
and are thus salient to attract gaze (Figure \ref{fig:V1Role}A).
Numerous V1SH predictions have been confirmed in monkey electrophysiology \cite{YanZhaopingLi2018}, 
neuroimaging, and behavior \cite{ZhaopingBook2014, ZhaopingNewFramework2019}.

Hence, V1 functions as a motor cortex for gaze control. It is a key member of an attentional selection network, 
guiding gaze exogenously (reflexively) with short latencies \cite{YanZhaopingLi2018,WesterbergEtAl2023}, whereas frontal and 
parietal areas control gaze and attention endogenously in a slower and 
sustained manner \cite{NakayamaMackeben1989, MullerRabbitt1989, BisleyGoldberg2010, 
ZhouDesimone2011, ZhaopingBook2014, KlinkEtAl2023}.  (The superior colliculus mediates most of 
the control signals to the brain stem.)
In vision research, exogenous selection has been relatively under-appreciated, likely 
due to its involuntary nature and our under-appreciation of the attentional bottleneck.

Since V1 guides gaze, the bottleneck should start immediately 
at V1's output to downstream visual stages to reduce cost, 
motivating the central-peripheral dichotomy (CPD) theory
\cite{ZhaopingNewFramework2019} which proposes the following:
(1) the peripheral and the central visual fields are specialized for looking and seeing,
and (2) because information flow through the bottleneck is limited, 
downstream stages send feedback to upstream stages to query for additional relevant information to 
aid ongoing recognition, and this feedback is directed mainly to the central visual field. 

Having the richest information upstream, V1 is a particular target of the feedback.
Hence, feedback to V1 is predicted to mainly target V1's central field representation (Figure \ref{fig:V1Role}A).
This prediction is supported by neuroimaging \cite{SimsEtAl2021} 
and neurophysiological \cite{MoralesEtAl2024} data, and
is confirmed by retrograde tracer injections to marmoset monkey V1, which
show an order-of-magnitude decrease in feedback prevalence with 
increasing visual-field eccentricities \cite{MajkaZhaopingRosa_VSS2026}.

Lacking the feedback query, the peripheral field is 
vulnerable to ambiguities (e.g., crowding) and illusions
 (e.g., the scintillating illusion, Figure \ref{fig:V1Role}B). 
In contrast, vision in the central field vetoes such illusions because 
the additional information queried by the feedback disambiguates perception.
This feedback query uses analysis-by-synthesis to 
test perceptual hypotheses, e.g., between the two competing hypotheses that the object is a red flower 
(hypothesis $H_1$) or a red apple (hypothesis $H_2$) 
in Figure \ref{fig:V1Role}A. 
It synthesizes would-be upstream responses for each hypothesis, and compares 
the would-be with the actual responses (queried by the feedback) 
to verify or disambiguate between alternative 
hypotheses \cite{ZhaopingNewFramework2019}.  
This predictive synthesis  makes, e.g., the Kanizsa triangle appear vivid (Figure \ref{fig:V1Role}B) \cite{MooreEtAl2025}.
The CPD theory predicted two new peripheral illusions: the flip tilt 
illusion ( Figure \ref{fig:V1Role}C) and the reversed depth illusion.  
Both illusions have been confirmed, and, as predicted, both become visible 
in the central visual field when the feedback is disrupted by backward 
masking \cite{ZhaopingFlipTilt2020, Zhaoping_iScience2025, Zhaoping_VSS2025}. 

The CPD theory offers  new perspectives. A retinotopic cortical area is not only 
positioned in a visual processing hierarchy, but also contains qualitatively different subregions 
for central versus peripheral functions.  Visual phenomena such as imagery, 
crowding,  amblyopia, and illusion can now be examined for links with a rich 
or absent feedback query \cite{ZhaopingNewFramework2019, ZhaopingPeripheral2024}.

\section*{Object based attention, transsaccadic integration, and working memory}

I conjecture that the feedback query through the bottleneck leads to
object based attention, even when this attention is task irrelevant \cite{EglyEtAl1994, Chen2012},
such as when one sees the object {\sf X} when only its component 
bars are task-relevant in Figure \ref{fig:LookingAndSeeing}A.  
A perceptual hypothesis $H = \{\theta_1, \theta_2, ...\}$ about an object 
can concern feature dimensions
such as orientation and color, with respective feature 
values such as $\theta_1$ and $\theta_2$.  
Processing one feature $\theta_i$ for $H = \{\theta_1, \theta_2, ... \}$ gives a 
processing advantage to another feature $\theta_j$ of the same, rather than a different, object.

When one feature $\theta_i$ is about the viewpoint of the object, 
its value  can be updated across a saccade using the efference copy of the 
saccadic command \cite{Wurtz2018}, thereby updating the would-be 
responses at upstream stages after the saccade.  This reduces the discrepancy 
between the expected and the actual upstream responses after the saccade, 
including for relevant or salient objects that are not the saccadic target,
thereby maintaining a stable perception.
Transsaccadic integration of visual inputs refines percepts 
\cite {Irwin1991, StewartSchuetz2018, LiangZhaoping2024} 
and manifests a phenomenological feedback query from a peripheral location before the saccade to 
a central location after the saccade \cite {WilliamsEtAl2008, FanEtAl2016, KnapenEtAl2016}. 

Compared to previous ideas of top-down predictive mechanisms for perceptual 
stability across saccades (e.g., \cite {Helmholtz1925, DayanEtAl1995, WangEtAlQian2024, CavanaghMelcher2026}),
our perspective in the looking-and-seeing framework 
emphasizes the bottleneck, so that the percept $H$ 
has less information than upstream visual signals.
Consequently, sufficiently small changes in a visual display (e.g., about the location or other features of a saccadic target)  
during saccades are unnoticeable \cite {Irwin1991}, since they cause little change in $H$. 
Although impoverished, scene information in the $H$'s can accumulate within a fixation and across fixations, 
 partly by iterating the queries 
(small-amplitude fixational eye movements can be non-random to serve the query \cite{IntoyRucci2020, WittenEtAl2024}), 
and be stored in capacity-limited visual working memory \cite{LuckVogel2013, VanDerStigchelHollingworth2018} 
or consolidated into longer-term memories.

\section*{Extensions and challenges}

\begin{figure}[h!!!]
\begin{center}
\includegraphics[width=150mm]{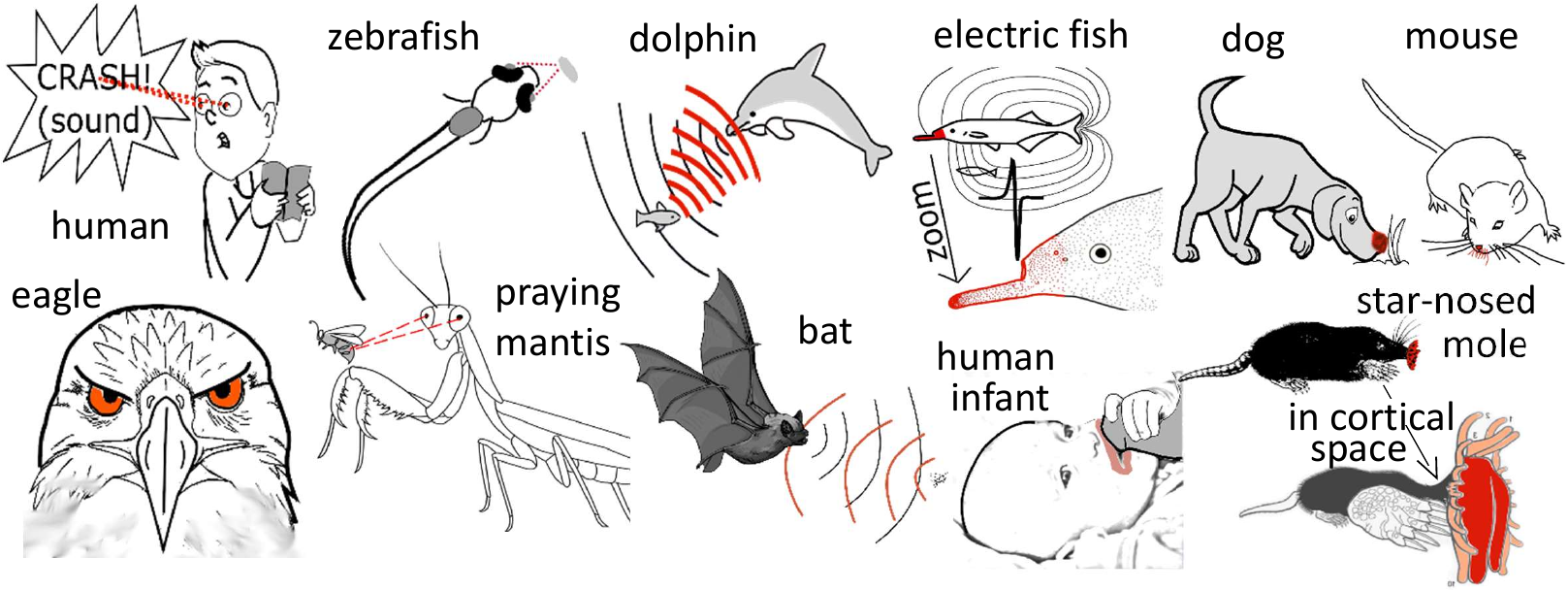}
\end{center}
\caption{\label{fig:MultiSensory}
Generalizing vision as looking and seeing to multisensory sensing as orienting (selection)
and recognition (decoding) across species and developmental stages in a central-peripheral dichotomy,
as all animals have a processing bottleneck.  In each of the eleven examples shown here,  
the sense organ specialized for the central function of recognition is indicated by a red hue.  In human adult, 
the peripheral function to select is served not only by vision in the peripheral field, 
but also by audition, touch, and smell to direct gaze shift, placing the selected (attended) object to the visual fovea 
for recognition and scrutiny through the bottleneck. 
Among the eleven examples, the central function to recognize is served by vision in human adult, eagle, zebrafish, and praying mantis, 
by audition in dolphin and bat, by electric sense in electric fish, and by touch and olfaction in human infant, dog, mouse, and 
star-nosed mole.  The central sense is apparent in the orienting behavior, and manifested in the underlying 
brain structure and neural physiology.  Figure adapted from \cite{ZhaopingCPD2023}.
}
\end{figure}

Many animal species have no visual fovea and limited eye movements, yet all
brains have a processing bottleneck.  Thus, looking and seeing can be generalized to multisensory 
orienting and recognition. 
The ``fovea" for scrutinization may be (e.g.,) olfactory, auditory, and/or somatosensory (Figure \ref{fig:MultiSensory}).
For example, rodent vision serves largely peripheral orienting, 
selecting sensory inputs into the bottleneck for recognition mainly by olfaction 
and microvibrissal touch \cite{BrechtEtAl1997, DiamondEtAl2008}. 
CPD provides a framework to link ecological, behavoiral, neurophysiological, 
and anatomical data, generating falsifiable predictions \cite{ZhaopingCPD2023}.

Vision can also select (look) and navigate, escape, avoid, and act with or without 
seeing \cite{GoodaleMilner1992, LandHayhoe2001, SerenoHuang2014, VaterETAl2022, ZhaopingPeripheral2024}.  
Thus, the looking-and-seeing framework for vision is an approximation, 
albeit an extension from vision as merely seeing.
It motivates free-viewing experimental paradigms 
that avoid enforced, non-natural fixations \cite{YatesEtAl2023}. 
It challenges us to understand extrastriate visual cortices accordingly.  
Building on confirmed predictions by V1SH and the CPD theory, we can proceed by 
testing those that remain untested while  deriving additional falsifiable predictions.

\section*{Acknowledgement}
This work is supported in part by the University of T\"ubingen and the Max Planck Society.
I thank Peter Dayan for comments on the manuscript.


\end{document}